\pdfoutput=1
\documentclass{PoS}

\usepackage{amsmath}
\usepackage{multicol}
\newcommand{\Tr}{{\rm Tr}}
\newcommand{\csw}{{c_{\mathrm{sw}}}}

\title{Determination of \(c_{\rm sw}\) in \(N_f=3+1\) Lattice QCD with massive Wilson fermions}

\ShortTitle{Determination of \(c_{\rm sw}\) in \(N_f=3+1\) Lattice QCD with massive Wilson fermions}

\author{\hfill\parbox{3cm}{\small\it%
DESY 15-015    \\
HU-EP-15/06    \\
SFB/CPP-14-116 \\
}}

\author{Patrick Fritzsch\\
        Humboldt-Universit\"at zu Berlin, Institut f\"ur Physik, Newtonstr. 15, 12489 Berlin, Germany \\
        E-mail: \email{fritzsch@physik.hu-berlin.de}}
        
\author{Rainer Sommer\\
        NIC, DESY, Platanenallee 6, 15738 Zeuthen, Germany \\
        E-mail: \email{rainer.sommer@desy.de}}

\author{\speaker{Felix Stollenwerk}
\\
        Humboldt-Universit\"at zu Berlin, Institut f\"ur Physik, Newtonstr. 15, 12489 Berlin, Germany \\
        E-mail: \email{felix.stollenwerk@physik.hu-berlin.de}}
        
\author{Ulli Wolff\\
        Humboldt-Universit\"at zu Berlin, Institut f\"ur Physik, Newtonstr. 15, 12489 Berlin, Germany \\
        E-mail: \email{uwolff@physik.hu-berlin.de}}
        
\abstract{
We develop a strategy for the non-perturbative determination of the \(O(a)\)-improvement coefficient \(c_{\rm sw}\) for Wilson fermions with massive sea quarks. The improvement condition is defined via the PCAC relation in the Schr\"odinger functional. It is imposed along a line of constant physics designed to be close to the correct mass of the charm quark. The numerical work uses the tree-level improved L\"uscher-Weisz gauge action in \(N_f=3+1\) Lattice QCD.
}

\FullConference{The 32nd International Symposium on Lattice Field Theory,\\
		23-28 June, 2014\\
		Columbia University New York, NY}

\begin{document}

\section{Introduction}

In Lattice QCD, the Wilson discretization of the fermion action introduces $O(a)$ cutoff effects, which may be systematically reduced by the well-established Symanzik improvement programme \cite{Symanzik}. In the case of massless fermions, only the clover term with the Sheikholeslami-Wohlert coefficient $\csw$ \cite{Sheikholeslami:1985ij} needs to be added to the action, but composite fields require additional improvement terms \cite{Luscher:1996sc}.   

The function $\csw(g_0^2)$ has been determined for various numbers of massless, dynamical quark flavours and different actions \cite{Luscher:1996ug, Jansen:1998mx, Yamada:2004ja, Aoki:2005et, Cundy:2009yy, Tekin:2009kq, Bulava:2013cta}. 
It has found successful application in simulations with
{\em massless} quarks (e.g. \cite{Tekin:2010mm} with $N_f=4$) or {\em light} quarks (e.g. \cite{Aoki:2009tf, Bruno:2014jqa} with $N_f=3$).
In the latter case, in addition to the clover term there exist {\em mass-dependent} improvement terms of order $O(a m_{q,\rm light})$, which can usually be treated perturbatively or even be neglected if $am_{q,\rm light}$ is small. 

The present work aims at improvement of the general, non-degenerate four flavor Wilson fermion action with a {\em massive} charm quark. In that case, an incomplete knowledge of the coefficients of the mass-dependent
improvement terms introduces large cutoff effects, see below.
We want to avoid both these large effects and the non-perturbative determination of a large number of improvement coefficients. 
This is possible by switching to a {\em massive} improvement and renormalization scheme, in which the improvement condition for $\csw$ is formulated at {\em finite} charm quark mass along a line of constant physics (LCP).

\section{Symanzik improvement with four non-degenerate quark flavors \label{sec:symanzik}}
We are interested in lattice QCD simulations with four flavors of Wilson quarks $\psi$,
\begin{align}\label{eq:wilson}
        \mathcal{L}       &= \mathcal{L}_{G} + \mathcal{L}_{F}   \;, & 
        \mathcal{L}_{F}   &= \bar{\psi} ( D_W + M + m_{\rm crit}) \psi         \;,  \\
        \mathcal{L}_{I,0} &= \mathcal{L} + a\mathcal{L}_{\rm sw}      \;, & 
        \mathcal{L}_{\rm sw} &= c_{\rm sw}(g_0^2)\bar{\psi} \tfrac{i}{4}\sigma_{\mu\nu}F_{\mu\nu} \psi \;,
\end{align}
and consider a non-degenerate matrix of bare subtracted quark masses 
\begin{align}\label{eq:Mmatrix}
          M &= {\rm diag} (m_{q,u}, m_{q,d}, m_{q,s}, m_{q,c}) \;, &
    m_{q,i} &= m_{i,0} - m_{\rm crit} \;.
\end{align}
The particular choice of a lattice gauge action $\mathcal{L}_G$ is irrelevant
for the present discussion, but will be the tree-level improved L\"uscher-Weisz
gauge action in section~\ref{sec:results}.  While $\mathcal{L}_{I,0}$ is the correct
on-shell $O(a)$ improved action density for simulations at the critical line
($m_{q,i}=0$), additional mass-dependent $O(a)$ terms arise for non-vanishing quark
masses. They read
\begin{align}\label{eq:LM}
      \mathcal{L}_{aM} &= \sigma_1         \bar{\psi}M^2\psi  
                        + \sigma_2 \Tr(M)  \bar{\psi}M  \psi
                        + \left( \sigma_3 \Tr(M^2) + \sigma_4 \Tr(M)^2 \right)\bar{\psi}\psi
                        + \sigma_5 \Tr(M)  \Tr(F_{\mu\nu}F_{\mu\nu})
\end{align}
and should be included in the simulation
($\mathcal{L}_{I,M}=\mathcal{L}_{I,0}+a\mathcal{L}_{aM}$) with
suitably chosen coefficients $\sigma_{i}(g_0^2)$. 
One may absorb the above improvement terms into a redefinition 
of the bare parameters of the theory, 
\begin{align}\label{eq:tilde} 
   g_0^2   &\to \tilde g_0^2   = g_0^2 ( 1 + a b_g(g_0^2) \Tr[M] / N_f) \;, &
   m_{q,i} &\to \tilde m_{q,i} = m_{q,i} f_i(g_0^2,\{aM\})   \;,
\end{align}
where $f$ is a non-trivial function of the bare subtracted quark masses in 
$aM$~\cite{Luscher:1996sc,Bhattacharya:2005rb} and a number of improvement 
coefficients related to those in~\eqref{eq:LM}.

For full $O(a)$ improvement of the action at finite quark
mass, a non-perturbative determination of the numerous improvement coefficients is in principle necessary. 
However, this is clearly out of reach for any number of flavors. 
In the past, without a charm quark, 
one could circumvent the problem due to the smallness of the mass-dependent improvement terms. For typical lattice spacings \mbox{$a \leq 0.1~{\rm fm}$}, assuming
\mbox{$m_{q,s} \approx 100~{\rm MeV}$} and \mbox{$m_{q,c} \approx 1~{\rm
GeV}$}, they are roughly of sizes 
\begin{equation}
        am_{q,s} \lesssim 0.05 \ , \qquad am_{q,c} \lesssim 0.5 \ .
        \label{eq:amqsc} 
\end{equation}
Hence, a viable way for small lattice spacings and quark masses, i.e.  simulations with $N_f\le3$, has been to treat the terms appearing in $\mathcal{L}_{aM}$ as small 
and account for them approximately. 

Regarding (\ref{eq:amqsc}), the effects of the charm mass in the improved theory are numerically much more important, which makes the non-perturbative treatment of all improvement coefficients unavoidable.
Even if such a treatment
was feasible, any uncertainties in the improvement coefficients would be amplified by the large charm mass, leading to a significant 
uncertainty in the improved theory.
Note that the same issue concerns the renormalization constants, for which it arises already at the $O(a^0)$ level,
\begin{eqnarray}
  m_{R,i} = Z_m &\bigg[& m_{q,i} + \left( r_m - 1 \right) \frac{\Tr[M]}{N_f} \bigg] + O(a) \ . 
 \label{eq:quarkmassesrenormalizedimproved}
\end{eqnarray}
In particular for observables in the light sector (no charm valence quarks involved), the large charm mass cutoff and renormalization effects and assigned uncertainties would be likely to surpass the small physical effects of a massive dynamical charm quark \cite{Bruno:2014ufa} that one is interested in. 

\section{Massive renormalization and improvement scheme \label{sec:massive_impr_renor}}

For the reasons discussed in the previous section, the massless renormalization and improvement scheme is impractical and unstable if a physical charm quark is to be included in the sea. 
Instead, we choose a massive scheme, in which the mass dependent 
improvement terms are absorbed in the bare parameters Eq.~(\ref{eq:tilde}),
which are tuned such that a LCP is followed. 
Accordingly, the renormalization and improvement conditions are imposed at finite quark masses, and the coefficients become mass-dependent:
\begin{equation}
 Z(g_0^2, a\mu) \to Z(g_0^2, aM, a\mu), \quad c(g_0^2) \to c(g_0^2, aM) \ . \label{eq:Zcmassive}
\end{equation}  
Here, $Z$ and $c$ are renormalization constants (depending in general on a
renormalization scale $\mu$) and improvement coefficients, respectively, and we recall that $aM$ is the bare subtracted mass matrix.
The action in the massive scheme then simply reads
\begin{equation}
 \label{eq:Smassive}
 \mathcal{L}_{I,M} = \mathcal{L} + a c_{\rm sw}(g_0^2, aM) \bar\psi {\frac{i}{4}} \sigma_{\mu\nu} F_{\mu\nu} \psi \ .
\end{equation}
This scheme dispenses with the need for the non-perturbative determination of a large number of improvement coefficients at the cost of a more
complicated renormalization pattern. 
Note that the mass-dependence of the improvement coefficients, Eq.~(\ref{eq:Zcmassive}), is not a necessity for $O(a)$ improvement, as the difference is of $O(a^2)$ in the action. However, we will use this form both for consistency and in the hope of reducing the overall $O(a^2)$ cutoff effects.

\section{Determination of $\csw$ in $N_f=3+1$ \label{sec:csw}}

There are two qualifications to be added to the determination of $\csw$ in the massive scheme.
First, the major effect of the non-vanishing masses on $\csw$ stems from the charm quark. In order to substantially reduce the effort, it is therefore reasonable to treat the up, down, and strange quark as mass-degenerate light quarks with mass $am_{q,l}$, i.e. to determine $\csw$ in $N_f=3+1$.

Second, it is not practicable to vary all the bare parameters ($g_0^2, am_{q,l}, am_{q,c}$) 
and find an interpolating formula for $\csw$ in this 3-dimensional space.
Hence, we will fix the masses to approximately physical values, indicated by an asterisk. These two approximations,
\begin{equation}
 \csw(g_0^2, aM)  \xrightarrow{N_f=3+1} \csw(g_0^2, am_{q,l}, am_{q,c}) 
                  \xrightarrow[\text{masses}]{\text{fix~ren.}} \csw(g_0^2, am_{q,l}^\star, am_{q,c}^\star) \ ,
 \label{eq:csw3+1approximations}
\end{equation}  
together explicitly introduce $O(a^2)$ effects proportional to
\begin{eqnarray}
 && m_{q,u} - m_{q,l}^\star, \quad m_{q,d} - m_{q,l}^\star, \quad m_{q,s} - m_{q,l}^\star, \qquad m_{q,c} - m_{q,c}^\star \ ,
 \label{eq:order_asq_effects}
\end{eqnarray} 
which modify the already existing cutoff effects of this order.
These modifications are small if the mass differences are small.
The relevant point obviously is to ensure that the last difference
is not too large.

\subsection{Line of constant physics \label{sec:LCP}}

The improvement condition on $\csw$ will be imposed with the help
of Schr\"odinger functional correlation functions (sec.~\ref{sec:impr_cond}).
The relevant physical scales of the $N_f=3+1$ theory in such a finite volume are the size $L$ of the box and the renormalization group invariant masses $M_l$ and $M_c$ (not to be confused with the mass matrix $M$) of the light and charm quark, respectively. 
The bare masses are fixed to $m_{q,l}^\star$ and $m_{q,c}^\star$ (cf. Eq.~(\ref{eq:csw3+1approximations})) by requiring the RGI masses to assume their physical values, $M_l^\star$ and $M_c^\star$. 
In order to fix these in physical units, a scale is needed. The obvious choice is $L$, which we keep constant as well, at a value $L^\star$ to be specified: 
\begin{equation}
 L = L^\star, \quad L M_l = L^\star M_l^\star, \quad L M_c = L^\star M_c^\star \ .
 \label{eq:LCP_fixed}
\end{equation} 
Not only is the approach to keep all physical scales fixed convenient, it also has the advantage of avoiding possibly large $O(a)$ ambiguities in $\csw$, by forcing these to vanish proportional to $a$
in the limit $g_0^2 \to 0$. A discussion 
can for instance be found in section I.2.4.1 of \cite{Sommer:2006sj}.
The conditions (\ref{eq:LCP_fixed}) define a line of constant physics (LCP), i.e. they determine the bare parameters $(g_0^2, am_{q,l}^\star, am_{q,c}^\star)$ for a given lattice resolution $L/a$. 

In order to realize the LCP on the lattice, it is reformulated in terms of three more easily accessible quantities, 
\begin{eqnarray}
  \Phi_1 &=& \bar g_{GF}^2(L) \\
  \label{eq:Phi1}
  \Phi_2 &=& L \cdot \Gamma_{ud} 
  \label{eq:Phi2} \\
  \Phi_3 &=& L \cdot \left( \Gamma_{uc} - \frac12 \Gamma_{ud} \right) 
  \label{eq:Phi3} \ .
\end{eqnarray} 
Here, $\bar g_{GF}^2$ is the gradient flow coupling \cite{Luscher:2010iy} as defined in \cite{Fritzsch:2013je} for Schr\"odinger functional boundary conditions with $c=0.3$ by means of the Wilson flow. The functions
\begin{equation}
   \Gamma_{ij} = - \tilde \partial_0 \log \left( f_{A,I}^{ij} (x_0) \right) \big|_{x_0=T/2} \ ,
   \label{eq:Gamma_ij}
\end{equation}
are effective pseudoscalar meson masses with quark flavors $i,j$ defined in terms of the Schr\"odinger functional correlation function $f_A$ (see e.g. Eq.~(6.11) in \cite{Luscher:1996sc} for its definition).
All $\Phi_i$ are associated with $T=L$ and vanishing boundary gauge fields.
While each of them is a function of all three physical scales (or equivalently all bare parameters), they were chosen such that each of them dominantly depends on a single parameter.  
In terms of the $\Phi_i$, Eq.~(\ref{eq:LCP_fixed}) gets replaced by
\begin{equation}
 \Phi_i = \Phi_i^\star \ , \qquad i=1,2,3 \ ,
 \label{eq:LCP_implicit}
\end{equation} 
where the (continuum) values $\Phi_i^\star$ correspond to the scales $L^\star, M_l^\star$ and $M_c^\star$. In order to do the translation between the two sets of quantities, one needs scale setting, mass renormalization and the input parameters $M_l^\star$ and $M_c^\star$. Since this is not available in $N_f=3+1$, we resort to $N_f=2$ results. While this alters the particular LCP that is established, it again only concerns the explicitly introduced $O(a^2)$ effects in (\ref{eq:order_asq_effects}) via the values with asterisk.
In $N_f=2$, the setting of the scale is available from \cite{Fritzsch:2012wq}, and the hopping parameters which correspond to the input parameters $M_l^\star$ \cite{Fritzsch:2012wq} and $M_c^\star$ \cite{Heitger:2013oaa} can be derived under use of mass renormalization constants \cite{Fritzsch:2012wq,Fritzsch:2010aw}, the improvement coefficient $b_m$ \cite{Fritzsch:2010aw} and the critical hopping parameter $\kappa_{\rm crit}$ \cite{Blossier:2012qu}.
We then reused the ensembles produced in the course of \cite{Blossier:2012qu} with 
\begin{equation}
 L^\star \approx 0.8~{\rm fm} \label{eq:Lstar}
\end{equation}
and $L^\star/a = 12,16,20,24,32$ in order to evaluate $\Phi_i^\star(a/L^\star)$. 
The results which we employ to define the LCP in its form (\ref{eq:LCP_implicit}) are the continuum extrapolated values. 
We find
\begin{equation}
 \Phi_1^\star = 7.31, \quad \Phi_2^\star = 0.59, \quad \Phi_3^\star = 5.96 \ ,
\end{equation}
from $L^\star \approx 0.8\,$fm, $M_l^\star \approx M_s^{(N_f=2)}/3 = 138/3~{\rm MeV}$ \cite{Fritzsch:2012wq}, and $M_c^\star \approx M_c^{(N_f=2)} = 1.51~{\rm GeV}$ \cite{Heitger:2013oaa}.

\subsection{Improvement condition \label{sec:impr_cond}}

Along the LCP, we impose the improvement condition for $\csw$ in the Schr\"odinger functional. As this follows a standard procedure, we will not cover it here and refer to the original literature \cite{Luscher:1996sc} instead. Concerning the choice of Schr\"odinger functional parameters, we follow \cite{Bulava:2013cta}, with only a few exceptions. 
First, with respect to the LCP, $T=L^\star$ is kept, while $L$ is reduced by a factor 2, $L=L^\star/2$.
This in particular implies that we use an even number of lattice sites in the temporal direction, in contrast to \cite{Bulava:2013cta}.
Furthermore, the improvement condition in our case is 
\begin{equation}
 \Delta M^{ud} = 0 \ .
 \label{eq:improvement_condition}
\end{equation}
The quantity $\Delta M^{ud}$ corresponds to $\Delta M$ (cf. Eq. (3.10) in \cite{Bulava:2013cta}) in the previous determinations of $\csw$ at vanishing quark masses. However, in contrast to these, we have to specify the quark
flavors (u,d) which enter, due to the different masses involved in our case. 
We thus use the PCAC relation in the light sector, as we are mainly interested in improving light quark physics. 
Moreover, we do not use a tree-level value $\Delta M^{(0)}$ on the right hand side of (\ref{eq:improvement_condition}), since this is not needed 
when one performs the improvement on a line of constant physics, $L=L^\star$.

\section{Results for $L/a=8$ \label{sec:results}}

\begin{figure}
 \centering
 \includegraphics[scale=0.5]{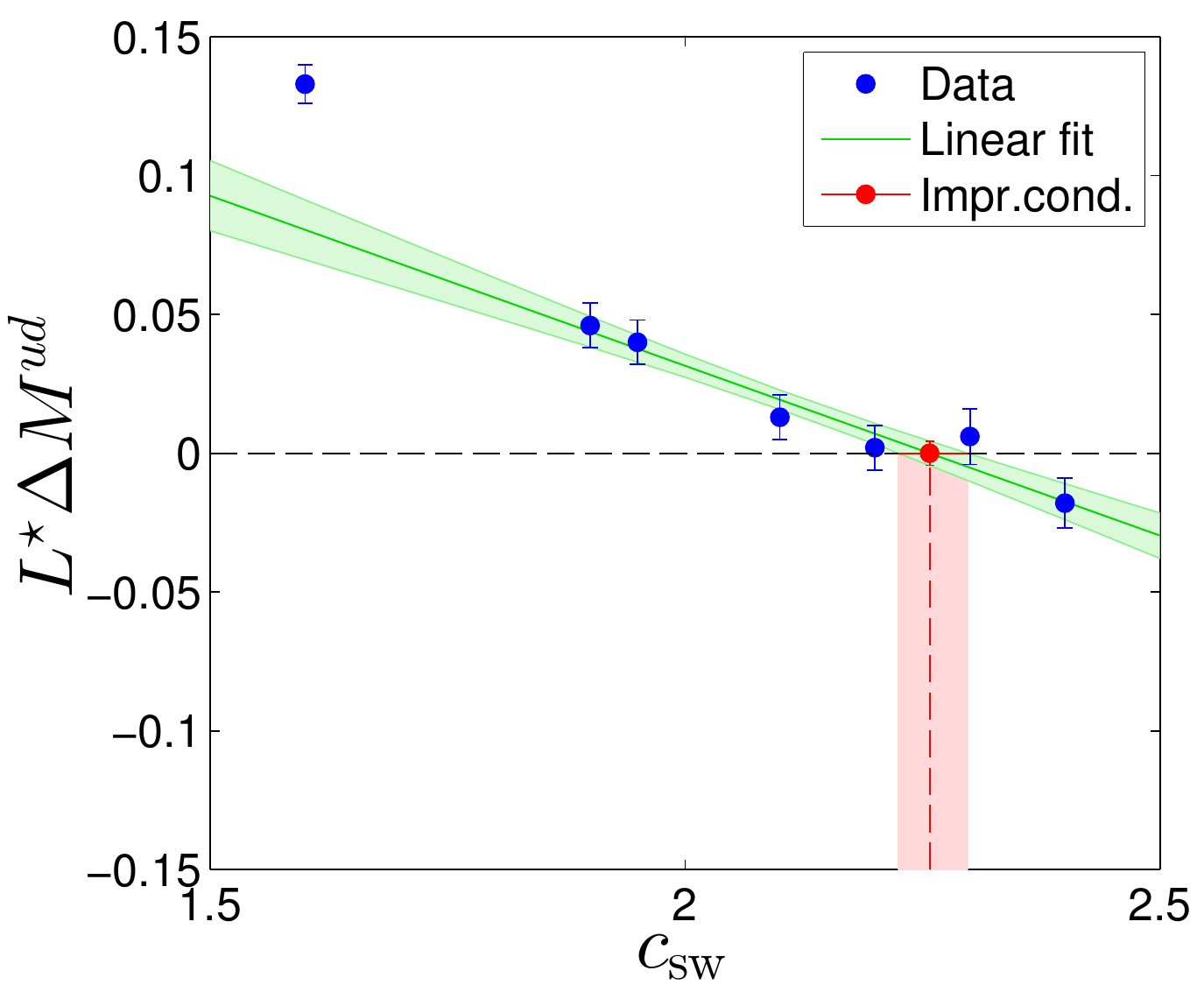}
 \includegraphics[scale=0.5]{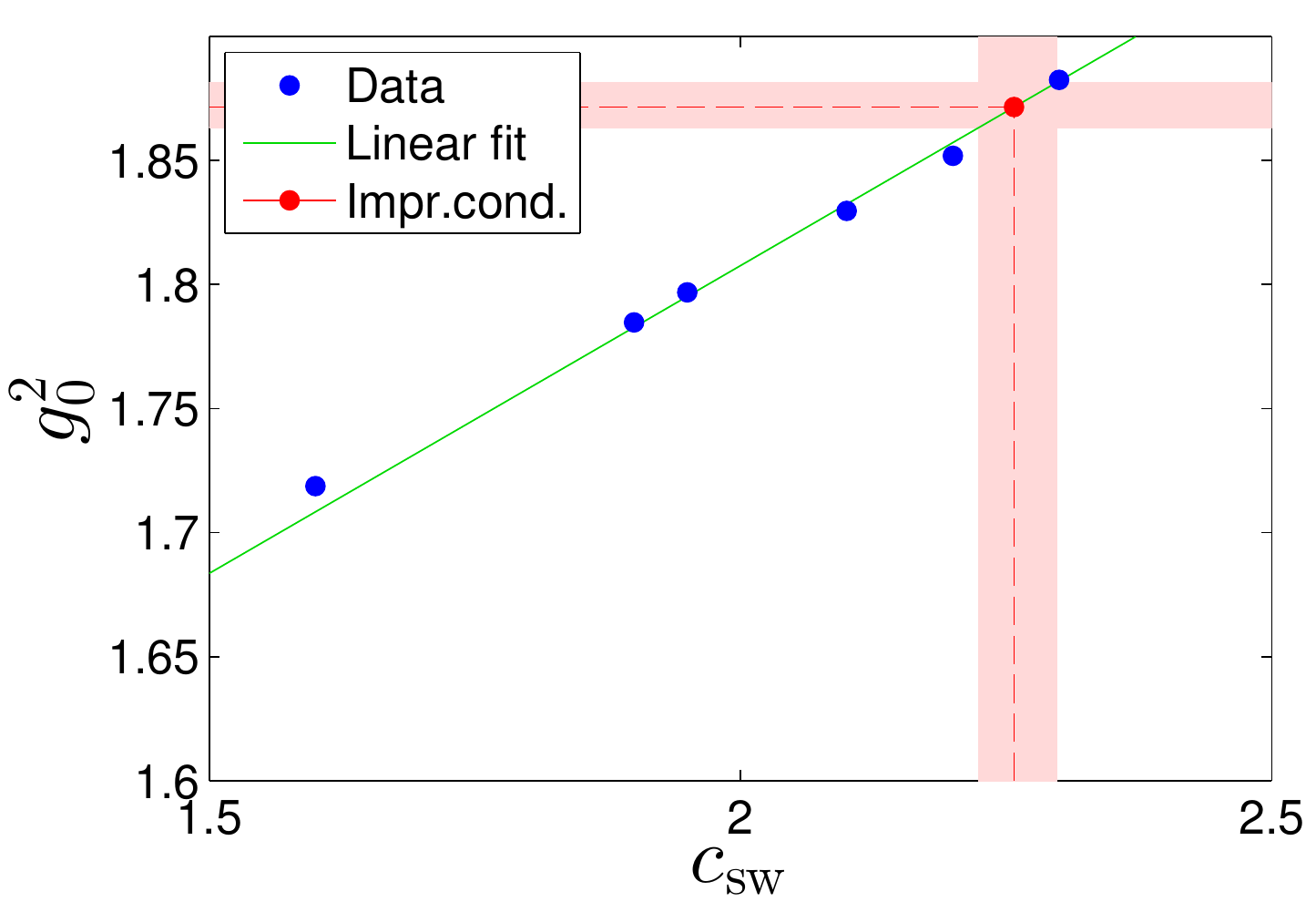}
 \caption{$L^\star \Delta M^{ud}$ (left panel) and $g_0^2$ (right panel) for $L/a=8$ as a function of $c_{\rm sw}$ and their linear interpolation (green) to the improvement point (red). The data point at $\csw=1.6$ is excluded from the fits as it is clearly outside the range of linear behaviour.}
 \label{fig:L8results}
\end{figure}

Preliminary results for $L/a=8$ are listed in Tab.~\ref{tab:L8results}.
\begin{table}
 \centering
 \small
 \begin{tabular}{l||c|c|c||c|c|c||c}
${c_{\rm sw}}$ & $g_0^2$ & $\kappa_l$ & $\kappa_c$ & $\Phi_1$ & $\Phi_2$ & $\Phi_3$ & $L^\star \Delta M^{ud}$ \\ \hline
$1.6$   & $1.7188$ & $0.1403$ & $0.1222$ & 
7.10(4)   & 0.56(2)    & 5.94(1) & $0.133(7)$  \\
$1.9$   & $1.7847$ & $0.1374$ & $0.1203$ &
7.18(2) & 0.59(2) & 5.96(1) & $0.046(8)$ \\
$1.95$  & $1.7969$ & $0.1369$ & $0.1199$ &
7.21(4) & 0.60(2) & 6.00(1) & $0.040(8)$ \\
$2.1$   & $1.8340$ & $0.1356$ & $0.1190$ & 
7.36(5) & 0.61(3) & 6.02(1) & $0.013(8)$ \\
$2.2$   & $1.8534$ & $0.1345$ & $0.1182$ & 
7.17(5) & 0.60(3) & 6.03(1) & $0.002(8)$ \\
$2.3$   & $1.8825$ & $0.13369$ & $0.1180$ &
$7.33(6)$ & $0.60(3)$ & $5.96(1)$ & $0.006(10)$  \\
$2.4$   & $1.9069$ & $0.13278$ & $0.1175$ & 
$7.34(6)$ & $0.58(4)$ & $5.94(1)$ & $-0.018(9)$
\end{tabular}
 \caption{Preliminary results for $L/a=8$. At the bare parameters given in columns 2-4, the values $\Phi_i$ sufficiently close to the LCP were found in $T=L=L^\star$ simulations with vanishing boundary fields. At the very same bare parameters, the quantity in the last column was obtained in simulations with half the spatial size and non-vanishing boundary fields. The simulations were performed with the HMC of the openQCD package \cite{luescher_openQCD} and error estimates have been obtained under use of the $\Gamma$-method \cite{Wolff:2003sm}.}
\label{tab:L8results}
\end{table}
The bare parameters were tuned such that the $\Phi_i$ 
satisfy the LCP conditions (\ref{eq:LCP_implicit}) with allowed deviations of up to 5\% for $\Phi_1,\Phi_3$ and 10\% for $\Phi_2$.  
The results for $L^\star \Delta M^{ud}$ as a function of $\csw$ are interpolated to the point where the improvement condition (\ref{eq:improvement_condition}) is fulfilled, as shown in the left panel of Fig.~\ref{fig:L8results}. A subsequent interpolation of $g_0^2$ to the previously obtained $\csw$ gives the corresponding $g_0^2$, see the right panel of Fig.~\ref{fig:L8results}. 
The results of this procedure are 
\begin{equation}
 \csw = 2.26(4), \qquad g_0^2 = 1.87(1). \label{eq:L8_results} 
\end{equation}

\section{Summary and outlook \label{sec:summary}}
We have developed a strategy to determine $\csw$ in $N_f=3+1$ with a massive charm quark. In the framework of a massive renormalization and improvement scheme, a line of constant physics is set up which aims at keeping in particular the charm quark mass fixed to an approximately physical value in order to avoid large mass-dependent cutoff effects.  
The first point for $\csw$ as a function of $g_0^2$ on the coarsest lattice with $L/a=8$, $a\approx 0.1$fm, has been computed. Simulations on finer lattices, which correspond to smaller $g_0^2$ and will enable contact with perturbation theory, are on the way. So far, the main effort was in tuning to the proper line of constant physics, but this step will become easier due to the gained knowledge. The invested effort is also valuable as the approximate knowledge of $\kappa_c(g_0^2)$ will help to plan subsequent simulations in the improved theory with a massive charm.

\end{document}